\documentclass[10pt,conference]{IEEEtran}

\usepackage{pdflscape}
\usepackage{booktabs}

\usepackage{cite}
\usepackage{amsmath,amssymb,amsfonts}
\usepackage{algorithmic}
\usepackage{graphicx}
\usepackage{textcomp}
\usepackage{xcolor}
\usepackage{subcaption}
\usepackage{hyperref}

\usepackage{courier}
\usepackage{listings}

\newboolean{showcomments}
\setboolean{showcomments}{true}
\ifthenelse{\boolean{showcomments}}
 { \newcommand{\mynote}[2]{
      \fbox{\bfseries\sffamily\scriptsize#1}      {\small$\blacktriangleright$\textsf{\emph{#2}}$\blacktriangleleft$}}}
        { \newcommand{\mynote}[2]{}}

\usepackage{booktabs}
\usepackage{multirow}
\begin{document}

\title{Are We SOLID Yet? An Empirical Study on Prompting LLMs to Detect Design Principle Violations}

\author{\IEEEauthorblockN{Fatih Pehlivan\IEEEauthorrefmark{1}, Arçin Ülkü Ergüzen\IEEEauthorrefmark{1}, Sahand Moslemi Yengejeh\IEEEauthorrefmark{1}, Mayasah Lami\IEEEauthorrefmark{1} and Anil Koyuncu\IEEEauthorrefmark{2}}
\IEEEauthorblockA{Bilkent University, Turkey \\ 
}
\IEEEauthorblockA{\IEEEauthorrefmark{1} \{fatih.pehlivan, ulku.erguzen, sahand.moslemi, m.lami\}@bilkent.edu.tr
}
\IEEEauthorblockA{\IEEEauthorrefmark{2} anil.koyuncu@cs.bilkent.edu.tr
} }

\maketitle
\begin{abstract}


Traditional static analysis methods struggle to detect semantic design flaws, such as violations of the SOLID principles, which require a strong understanding of object-oriented design patterns and principles. Existing solutions typically focus on individual SOLID principles or specific programming languages, leaving a gap in the ability to detect violations across all five principles in multi-language codebases. This paper presents a new approach: a methodology that leverages tailored prompt engineering to assess LLMs on their ability to detect SOLID violations across multiple languages. We present a benchmark of four leading LLMs—CodeLlama:70B, DeepSeekCoder:33B, Qwen2.5 Coder:32B, and GPT-4o Mini—on their ability to detect violations of all five SOLID principles. For this evaluation, we construct a new benchmark dataset of 240 manually validated code examples. Using this dataset, we test four distinct prompt strategies inspired by established zero-shot, few-shot, and chain-of-thought techniques to systematically measure their impact on detection accuracy. Our emerging results reveal a stark hierarchy among models, with GPT-4o Mini decisively outperforming others, yet even it struggles with challenging principles like DIP. Crucially, we show that prompt strategy has a dramatic impact, but no single strategy is universally best; for instance, a deliberative ENSEMBLE prompt excels at OCP detection while a hint-based EXAMPLE prompt is superior for DIP violations. Across all experiments, detection accuracy is heavily influenced by language characteristics and degrades sharply with increasing code complexity. These initial findings demonstrate that effective, AI-driven design analysis requires not a single ``best'' model, but a tailored approach that matches the right model and prompt to the specific design context, highlighting the potential of LLMs to support maintainability through AI-assisted code analysis.

\end{abstract}



\section{Introduction}

Ensuring high-quality, maintainable, and extensible software is a fundamental challenge in software engineering. While the SOLID principles—Single Responsibility (SRP), Open/Closed (OCP), Liskov Substitution (LSP), Interface Segregation (ISP), and Dependency Inversion (DIP)— provide a robust foundation for good design~\cite{ampatzoglou2019applying}, violations are common and degrade code quality~\cite{turan2018experimental}. LLMs are now being integrated into developer workflows, but a critical, unaddressed question remains: Do these models understand the principles of good software design, or are they architecturally naive? An LLM that generates functionally correct but architecturally flawed code poses a significant long-term risk to software maintainability.

Existing analysis methods are insufficient to answer this question. Traditional methods, such as AST-based static analysis for the OCP~\cite{royAutomatedVerificationOpen2024}, are often narrowly focused on a single principle. Manual case studies underscore the industrial relevance of SOLID~\cite{girjoabaRefactoringLegacyCode2024, yanakiev2025applying} but lack automation. The application of LLMs is particularly promising, given their demonstrated ability to identify related issues like code smells\cite{sadik2025benchmarking}. However, direct applications to SOLID principles remain preliminary, largely consisting of theoretical proposals rather than empirical benchmarks~\cite{martinsUseLargeLanguage2024}. 



This reveals a critical gap between conceptual proposals and practical validation: there is no systematic, empirical benchmark of LLM performance on SOLID violation detection, evaluated across a spectrum of models, programming languages, and prompting strategies. Progress is further hampered by the absence of a standardized, public benchmark dataset designed for this specific, multi-language problem.

This paper directly addresses these gaps. We present the first systematic evaluation of four state-of-the-art LLMs—\texttt{CodeLlama:70b}~\cite{rozière2024codellamaopenfoundation}, \texttt{DeepSeekCoder:33b}~\cite{guo2024deepseekcoderlargelanguagemodel}, \texttt{Qwen2.5-Coder:32b}~\cite{qwen2025qwen25technicalreport}, and \texttt{GPT-4o-mini}~\cite{openai2024gpt4o}—on their ability to detect SOLID violations across Python, Java, C\#, and Kotlin. To enable this evaluation, we introduce a new benchmark dataset and test four custom prompt strategies inspired by established zero-shot, few-shot, and reasoning-based techniques. Our goal is to provide a foundational understanding of LLM capabilities and limitations in this domain.

Our key contributions are:
\begin{enumerate}
\item A systematic benchmark of LLMs for SOLID violation detection, representing the first empirical study to assess performance across four distinct models, four programming languages, and four tailored prompt strategies.
\item A benchmark dataset of 240 manually validated code snippets for SOLID violation detection, covering all five principles at three difficulty levels and uniquely providing both violating and refactored code versions.
\item An open-source replication package containing all data, prompts, evaluation scripts, and raw model outputs to ensure full reproducibility, publicly available at: \small \url{https://doi.org/10.5281/zenodo.17008546}\
\end{enumerate}

\section{Related Work}\label{sec:related-work}

Previous works link the adherence of SOLID principles with maintainability, where violations correlate with more code smells and reduced quality. Ampatzoglou et al.\cite{ampatzoglou2019applying} show SRP violations raise common code smells, while Turan and Tanrıöver\cite{turan2018experimental} quantify maintainability gains (e.g., analyzability) from adhering to SOLID guidelines; and Yanakiev et al.\cite{yanakiev2025applying} demonstrate this by improving a legacy C++ system via DIP refactoring. A complementary thread targets design-time validation. Oktafiani and Hendradjaya\cite{oktafianiSoftwareMetricsProposal2018} propose class-diagram compliance metrics, while Chebanyuk and Markov~\cite{chebanyukApproachClassDiagrams2016} offer logic-based verification for class diagrams. While this thread confirms the importance of SOLID, its methods are either manual or confined to design-time artifacts, not automated, code-level detection.

A second thread focuses on automating code quality checks. Traditional static analysis tools such as SonarQube~\cite{sonarqube} and Codacy~\cite{codacy} detect general quality issues but are known for high false positive/negative rates~\cite{bessey2010few,johnson2013don}. More recent tools like DeepCode~\cite{snykcode} and Amazon CodeGuru~\cite{codeguru} leverage machine learning to mitigate these issues. Even advanced approaches like Intelligent Code Analysis Agents (ICAA) that combine LLMs with static analysis\cite{fan2023static} focus on general-purpose analysis. Thus, this thread lacks tools with a comprehensive focus on detecting violations across all five SOLID principles directly from code.

The most recent thread explores LLMs for understanding high-level design. The potential is clear, as prompt engineering techniques like few-shot, chain-of-thought, and role-based prompting have proven effective in related tasks like code smell detection~\cite{sadik2025benchmarking,wu2024ismell,white2024chatgpt}. However, direct applications of LLMs to SOLID principles remain preliminary. For instance, Martins et al.\cite{martinsUseLargeLanguage2024} proposed a GPT-4-based GitHub bot for code reviews, but their work is a theoretical proposal without quantitative evaluation. Other studies highlight SOLID's importance for developer understanding in ML code\cite{cabralInvestigatingImpactSOLID2024} or note its underutilization in ML pipelines~\cite{lopezInsightsUseSoftware2025}. Crucially, these studies underscore the relevance of SOLID but stop short of providing a systematic, empirical benchmark to evaluate LLM performance on this detection task.



In contrast to prior work, our study provides the first systematic evaluation of multiple LLMs across all five SOLID principles and four programming languages using a dedicated benchmark dataset. Unlike design-time metrics or narrowly scoped tools, we evaluate detection directly on code, using accuracy and F1 scores. By analyzing the impact of prompt strategies, we offer new perspective into how LLMs internalize software design principles, positioning them as reflective tools for assessing code quality.

\vspace{-0.8em}
\section{Methodology}\label{sec:methodology}

\begin{figure}
    \centering
    \includegraphics[width=0.99\linewidth]{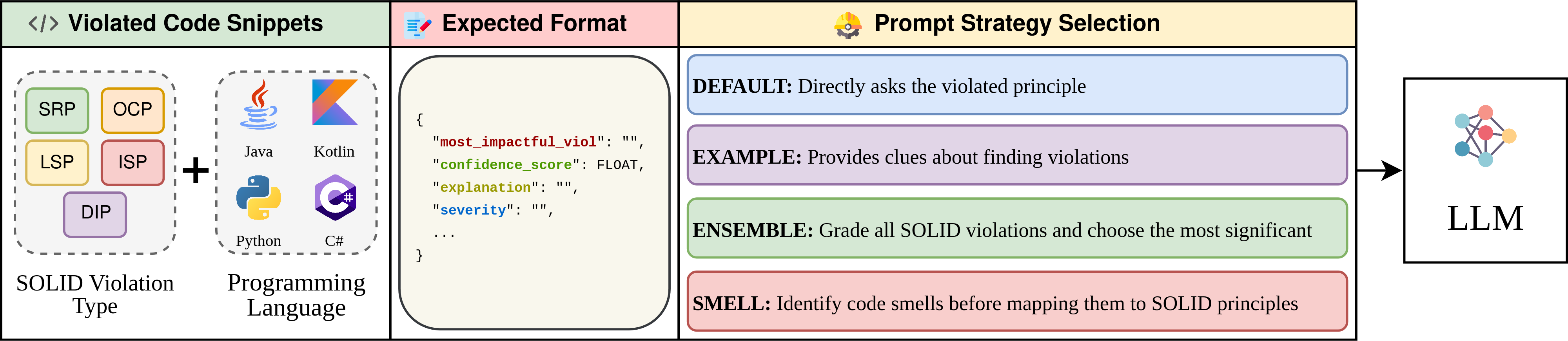}
\caption{Overview of the approach}
    \label{fig:problem_structure}
\end{figure}

Our methodology consists of three parts: (1) dataset creation, (2) prompt strategy design, and (3) model-based classification. Figure~\ref{fig:problem_structure} provides an overview of this approach, where each code snippet from our constructed dataset is analyzed by an LLM via a tailored prompt to predict a potential SOLID violation.


\vspace{-0.8em}

\subsection{Dataset Construction and Validation}
A key challenge in this domain is the absence of a public benchmark for SOLID violations. To address this, we construct a new dataset covering all five principles across four languages: \textbf{Java, Python, Kotlin, and C\#}.

Our creation process follows a hybrid methodology that combines LLM-based generation with manual authoring. First, we define 20 representative violation scenarios, four per principle, inspired by canonical patterns described in foundational software engineering literature~\cite{martin2000design}. For each scenario, we prompt OpenAI’s \texttt{gpt-4o} with structured requests specifying the violation type, target language, and complexity level. One author refined outputs for realism and clarity, implementing corresponding non-violating versions. Another author then revised the output as needed and implemented the corresponding non-violating version. A second author independently verified both versions for correctness and consistency. In case of disagreements, the first and second authors discussed and resolved conflicts through consensus.

\begin{figure}
    \centering
    \includegraphics[width=1\linewidth]{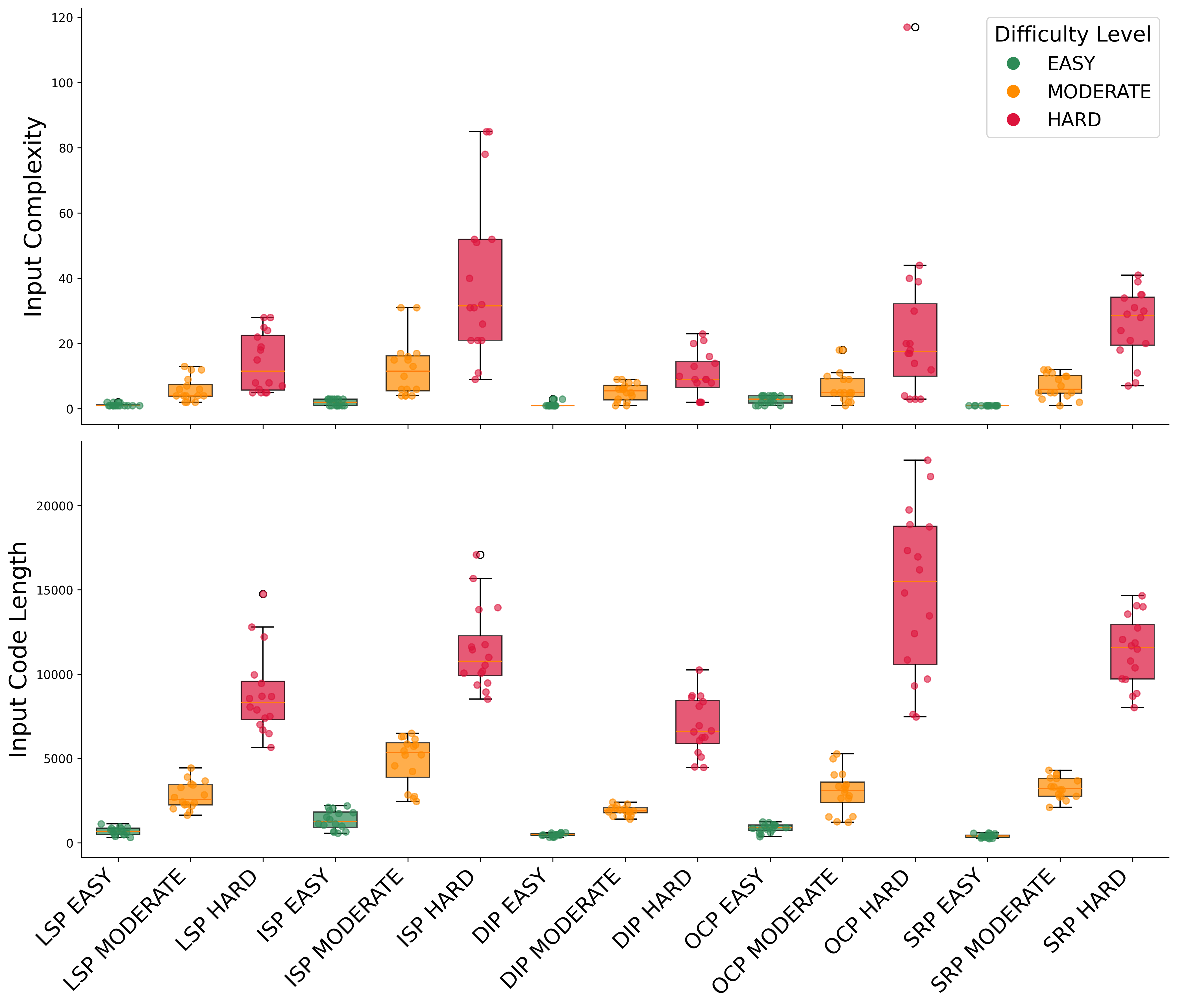}
    \caption{Code characteristics (character count, cyclomatic complexity) by SOLID violation and difficulty.}
    \label{fig:code_characteristics}
\end{figure}

To control for complexity, we implement each scenario at three difficulty levels: easy, moderate, and hard. We use character count an cyclomatic complexity as the proxy for difficulty. As Figure~\ref{fig:code_characteristics} shows, our assigned difficulty labels correlate strongly with both character count and cyclomatic complexity, validating our choice of proxy. The final dataset consists of 240 unique, manually validated code samples. A complete list of all 20 scenarios is available in our public replication package.

\subsection{Prompt Engineering Strategies}

We design and evaluate four prompt strategies, each requiring structured JSON output that classifies code into one of six classes (SRP, OCP, LSP, ISP, DIP, or No Violation) with a brief explanation. The baseline DEFAULT prompt is a direct zero-shot request~\cite{kojima2022large}. The EXAMPLE prompt uses a few-shot approach~\cite{brown2020language}, embedding one concise line per principle in the prompt to illustrate its violation (e.g., SRP: “unrelated responsibilities,” OCP: “repeated if/switch where polymorphism fits,” LSP: “subclass breaks base contract,” ISP: “fat interfaces,” DIP: “depends on concretes”). The SMELL prompt applies a two-step Chain-of-Thought process~\cite{wei2022chain} where the model is first asked to identify design smells, then map them to SOLID principles. The model then scores each principle (0–5), before outputting only the most violated one. Importantly, no explicit mapping between smells and principles is given, requiring inference from the models' prior knowledge. Finally, the ENSEMBLE strategy asks the model to score all five principles (0–5) with one-line justifications, then select and justify the single most impactful violation. \footnote{Full prompt templates and scripts are provided in the replication package.}

\subsection{Classification Process}\label{sec:evaluation}

To evaluate model performance, we process each model output against our dataset's ground-truth labels. We first attempt to automate this classification using tailored regular expressions designed to parse the unique output format of each prompt strategy. However, this automated approach reveals a significant challenge: LLMs frequently fail to adhere to the requested output structure. 

This widespread non-adherence requires us to manually review and label 1,431 out of 3,840 total responses (37\%). We perform this manual validation to resolve specific failure cases, such as when models detect multiple distinct violations, the regex fails to find a clear indicator, or language-specific response patterns emerge. This rigorous two-stage process ensures the quality and reliability of the final labels used for our analysis. \footnote{The specific regular expressions and detailed labeling criteria are available in our public replication package.}

\vspace{-1.5em}
\subsection{Evaluation Metrics}\label{sec:metrics}
To measure model performance, we use two standard classification metrics: Accuracy and F1-Score.


\vspace{-1em}

\begin{align}
\tiny \text{F1-Score} = \frac{2 \times \text{TP}}{2 \times \text{TP} + \text{FP} + \text{FN}}, \quad \text{Accuracy} = \frac{\text{TP} + \text{TN}}{\text{TP} + \text{FP} + \text{FN} + \text{TN}} \label{eq:metrics}
\end{align}

where $TP$, $FP$, $FN$, and $TN$ represent true positives, false positives, false negatives, and true negatives respectively.



\begin{figure}
    \centering
    \includegraphics[width=0.99\linewidth]{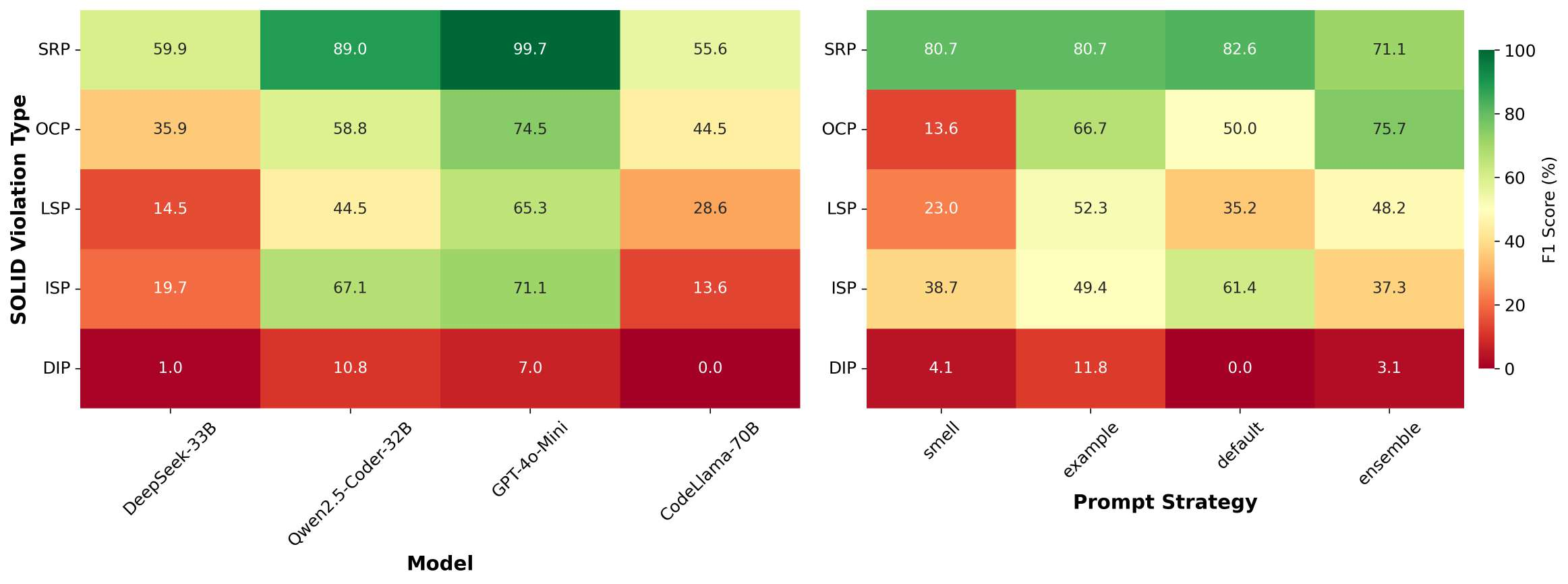}
    \caption{F1 scores of SOLID violation detection across a) LLM models and b) prompt strategies.}
    \label{fig:F1Scores}
\end{figure}

\begin{figure}
    \centering
    \includegraphics[width=0.99\linewidth]{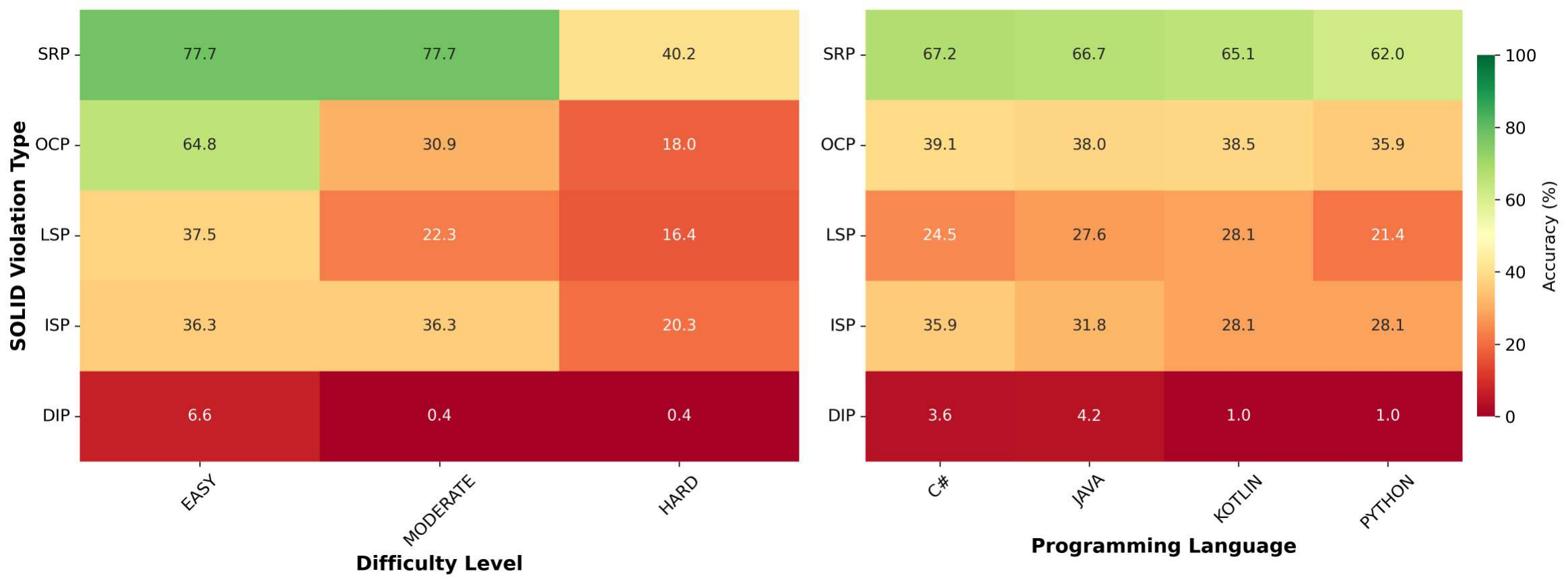}
    \caption{Average detection accuracy for a) each sample level and b) across programming languages per SOLID violation.}
    \label{fig:DetAccuracy}
\end{figure}

\vspace{-1.2em}
\subsection{Experiment Setup}
We design our experiments to answer our research questions by systematically evaluating four LLMs~\cite{rozière2024codellamaopenfoundation,guo2024deepseekcoderlargelanguagemodel,qwen2025qwen25technicalreport,openai2024gpt4o}, chosen to represent a diverse range of architectures, organizations, and sizes. All experiments are conducted at a temperature of 0 to ensure deterministic and reproducible outputs. We address each research question as follows:
\textbf{RQ1} investigates detection performance by comparing the accuracy of different LLMs across all violation types. \textbf{RQ2} assesses the impact of prompt engineering strategies by measuring changes in detection success rates for each SOLID principle under varying prompt formulations. \textbf{RQ3} evaluates language-specific performance.


\section{Evaluation}\label{sec:discussion}
\subsection{\textbf{RQ1:} What is the relative performance of different LLMs in detecting SOLID violations?}



Our results show a stark performance hierarchy among models. GPT-4o Mini is decisively the top performer, while other models struggle significantly, especially with more complex principles. As shown in Figure~\ref{fig:F1Scores}a, GPT-4o Mini demonstrates superior performance across nearly all principles, achieving exceptional F1-scores for SRP (99.7), OCP (74.5), and ISP (71.1). Qwen2.5-Coder-32B is a distant but clear second, showing competence in detecting SRP (89.0) and OCP (58.8) but faltering on more nuanced principles like DIP (10.8). The other models perform poorly; CodeLlama-70B's performance is weak outside of SRP (55.6), with a particularly low score for ISP (13.6), and DeepSeek-33B fails to achieve an F1-score above 40 for any principle other than SRP. For DIP, the most challenging principle, three of the four models are effectively unable to provide useful detections.

Furthermore, code complexity is a major factor. Figure~\ref{fig:DetAccuracy}a reveals a sharp decline in detection accuracy for all models when moving from easy to moderate and hard samples. While SRP remains relatively easy to detect regardless of complexity, accuracy on principles like DIP declines sharply, underscoring the challenge that complex code poses to current models.

\vspace{-0.35em}

\subsection{\textbf{RQ2:} How do different prompt strategies affect the ability of LLMs to detect violations of the SOLID principles?}
\vspace{-0.2em}
Prompting strategy is a critical factor, but no single strategy excels at all tasks. The ENSEMBLE and EXAMPLE strategies show strong, complementary performance, while the SMELL strategy is a consistent failure.

Figure~\ref{fig:F1Scores}b reveals a complex relationship between prompts and principles. The ENSEMBLE strategy is surprisingly effective for OCP (F1-score of 75.7), significantly outperforming all others. The EXAMPLE strategy proves most effective for LSP (52.3) and DIP (11.8), demonstrating that providing a hint is crucial for these nuanced violations. The baseline DEFAULT strategy excels at detecting SRP (82.6) and ISP (61.4), suggesting that for principles with clear structural patterns, a direct prompt is sufficient. In stark contrast, the SMELL strategy consistently underperforms, with catastrophically low scores for OCP (13.6) and LSP (23.0), suggesting that the indirect, two-step reasoning is ineffective for this task.

\vspace{-0.35em}
\subsection{\textbf{RQ3:} How does programming language affect the detection accuracy of SOLID violations?}
\vspace{-0.5em}
Detection accuracy is highly dependent on the programming language, with the structural clarity of statically-typed languages like C\# and Java leading to better performance, especially for simpler principles.

Figure~\ref{fig:DetAccuracy}b shows that C\# and Java yield the highest overall accuracy. For the most easily detected principle, SRP, they achieve scores of 67.2 and 66.7, respectively. This suggests that features common to these languages, such as explicit type declarations and formal class structures, provide clearer signals for LLMs. Kotlin follows, performing on par with Java for LSP (28.1 vs 27.6) but lagging elsewhere. Python, with its dynamic typing, consistently presents the greatest challenge, showing the lowest accuracy for four out of the five principles. This indicates that its syntactic flexibility makes design violations more ambiguous for automated tools.

\vspace{-0.8em}

\subsection{Cross-Cutting Finding: The Impact of Code Complexity}
\vspace{-0.2em}
Across all models, prompts, and languages, increasing code complexity is the greatest factor that degrades detection performance.

Figure~\ref{fig:DetAccuracy}a illustrates a sharp, universal decline in accuracy as samples move from EASY to MODERATE and HARD. For instance, OCP detection accuracy plummets from 64.8 on easy samples to just 18.0 on hard ones. This trend is even more pronounced for the most difficult principles; both LSP and DIP have accuracy scores below 25 for moderate and hard samples. While SRP detection remains somewhat robust, the overall pattern confirms that the selected LLMs struggle significantly to untangle design violations from general code complexity.

\vspace{-0.1em}
\textbf{Why models fail:}
We observed three recurring failures: (i) \emph{Principle ambiguity:} DIP and LSP require reasoning about abstractions that are harder to infer from code snippets. As a result, models tend to over-rely on more surface-level structural cues, leading to inflated detection of SRP and ISP violations. (ii) \emph{Two-step prompting:} The \textsc{SMELL} prompt requires the model to implicitly map design smells to SOLID principles without explicit guidance. This increases cognitive load and error propagation, contributing to its consistently low F1 scores.  (iii) \emph{Schema non-adherence:} Models frequently produce outputs that deviate from the expected format, necessitating manual review in 37\% of cases. Additionally, we observe a sharp decline in performance as code complexity increases, suggesting that incidental complexity can obscure the design-relevant signals the models are intended to detect.

\vspace{-0.8em}
\section{Conclusion and Future Work}\label{sec:conclusion}

We presented the first systematic evaluation of LLMs for detecting SOLID design principle violations across four models, languages, and prompt strategies on a new, manually validated dataset. Our findings provides emerging evidence that LLM effectiveness critically depends on the model, prompt, and code context. 

GPT-4o Mini emerged as the top performer, while direct, context-aware prompts (e.g., EXAMPLE) significantly outperformed abstract reasoning strategies. Statically-typed languages (C\# and Java) facilitated more accurate detection than dynamically-typed languages like Python.

This work has direct implications for how we assess AI coding assistants. An LLM's ability to reason about SOLID principles serves as a crucial proxy for its underlying ``design awareness.'' This proxy is vital because models lacking a grasp of these principles are likely to generate code that, while functional, degrades into less maintainable and extensible systems over time.


While this study provides insightful information, we acknowledge its limitations. Our findings are based on a synthetic dataset, which may not fully represent the complexity of industrial codebases. The results are a snapshot in time; the rapidly evolving LLM landscape may alter specific model rankings. Finally, the results may not generalize beyond the specific models, languages, and violation patterns we investigated. These limitations motivate several avenues for future work.


A key next step is to move from violation detection to automated refactoring. Future studies should task LLMs with generating corrections for the violations in our dataset. The quality of these LLM-generated solutions could then be rigorously assessed through a dual-evaluation approach: qualitatively via expert review against our manually written solutions, and quantitatively via automated test cases to verify that functional correctness is preserved. Further research should also expand this benchmark to include more models, real-world industrial code, and a wider range of design patterns.

\vspace{-0.5em}
\section{Acknowledgement}
We thank Rafi Çoktalaş and Muammer Buğra Kurnaz for their contributions to the initial prototype and discussions.

\bibliographystyle{IEEEtran}
\bibliography{sample-base}




\end{document}